\documentclass[sigconf]{acmart}
\AtBeginDocument{%
  }

\usepackage{subcaption}
\usepackage{wrapfig}
\usepackage{multicol,multirow}
\usepackage{float}
\usepackage{nicefrac}

\copyrightyear{2024}
\acmYear{2024}
\setcopyright{rightsretained}
\acmConference[RecSys '24]{18th ACM Conference on Recommender Systems}{October 14--18, 2024}{Bari, Italy}
\acmBooktitle{18th ACM Conference on Recommender Systems (RecSys '24), October 14--18, 2024, Bari, Italy}\acmDOI{10.1145/3640457.3688053}
\acmISBN{979-8-4007-0505-2/24/10}

\begin{document}

\title{Taming the One-Epoch Phenomenon in Online Recommendation System by Two-stage Contrastive ID Pre-training}

\author{Yi-Ping Hsu}
\email{yshu@pinterest.com}
\affiliation{%
  \institution{Pinterest Inc.}
  \country{USA}
}
\author{Po-Wei Wang}
\email{poweiwang@pinterest.com}
\affiliation{%
  \institution{Pinterest Inc.}
  \country{USA}
}
\author{Chantat Eksombatchai}
\email{pong@pinterest.com}
\affiliation{%
  \institution{Pinterest Inc.}
  \country{USA}
}
\author{Jiajing Xu}
\email{jiajing@Pinterest.com}
\affiliation{%
  \institution{Pinterest Inc.}
  \country{USA}
}

\renewcommand{\shortauthors}{Hsu et al.}

\newcommand*\jiajing[1]{\textcolor{red!50}{\textbf{jiajing: #1}}}

\newcommand*\yhsu[1]{\textcolor{red!50}{\textbf{yhsu: #1}}}

\newcommand*\powei[1]{\textcolor{red!50}{\textbf{powei: #1}}}

\begin{abstract}
 ID-based embeddings are widely used in web-scale online recommendation systems. However, their susceptibility to overfitting, particularly due to the long-tail nature of data distributions, often limits training to a single epoch, a phenomenon known as the "one-epoch problem." This challenge has driven research efforts to optimize performance within the first epoch by enhancing convergence speed or feature sparsity. In this study, we introduce a novel two-stage training strategy that incorporates a pre-training phase using a minimal model with contrastive loss, enabling broader data coverage for the embedding system. Our offline experiments demonstrate that multi-epoch training during the pre-training phase does not lead to overfitting, and the resulting embeddings improve online generalization when fine-tuned for more complex downstream recommendation tasks. We deployed the proposed system in live traffic at Pinterest, achieving significant site-wide engagement gains.
\end{abstract}


\maketitle

\section{Introduction}

ID-based embeddings have become essential in online recommendation systems, powering ranking models across  various social media platforms \cite{naumov2019deep,yin2021tt, covington2016deep,singh2023better, liu2022monolith, pi2019practice,chen2022efficient}. 

However, these systems face the challenge of the one-epoch overfitting phenomenon \cite{zhang2022towards,zhou2018deep,zhu2021open,liu2022monolith}, where the model overfits after just one pass through the data. This overfitting is caused by the power-law distribution  \cite{faloutsos1999power,leskovec2005graphs}, where tail entries possess more dimensions of freedom than training samples. Consequently, ID-based systems must optimize performance within the first epoch, often employing faster optimization techniques \cite{liu2022monolith} or thresholding \cite{zhang2022towards}.

Through experiments on web-scale recommendation systems, we observed that complex deep click-through rate (CTR) models without ID embeddings often require multiple epochs to converge. However, incorporating ID embeddings introduces the one-epoch phenomenon, resulting in suboptimal performance compared to multi-epoch models without ID-based embeddings. We propose a two-stage training approach that pre-trains ID embeddings in a minimal model using contrastive loss to address this issue. This method enhances data coverage and prevents overfitting during pre-training, thereby improving generalization and enabling multi-epoch training in downstream recommendation models. Our experiments demonstrate that this approach enables multi-epoch embedding training and enhances generalization when fine-tuning embeddings in downstream tasks. Additionally, online evaluations of our two-stage pre-training system show a 2.2\% increase in site-wide engagement.

\begin{figure*}%
    \centering
    \begin{minipage}[b]{0.58\textwidth}
        \centering
        \includegraphics[width=\textwidth]{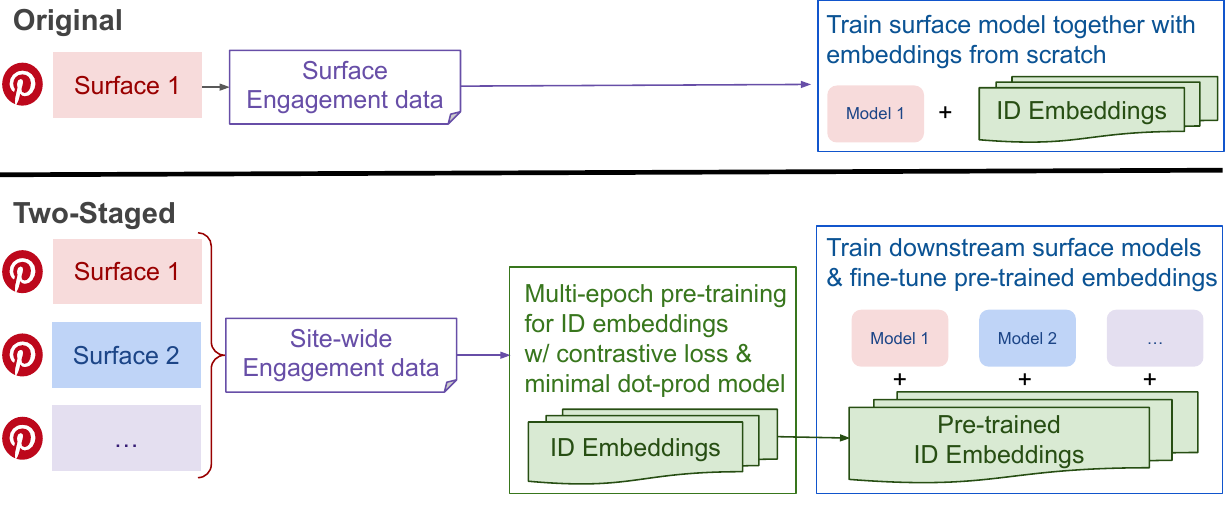}
        \vspace*{-2\baselineskip}
        \caption{The flow diagram for our proposed two-stage training strategy.  In Stage 1, we aggregated engagement data from multiple surfaces to pre-train the foundational ID embeddings in multiple epochs in a minimal contrastive learning model. Negative samples are added to the contrastive loss to reduce the effective dimension of freedom for tail entries. In Stage 2, the pre-trained ID embeddings are shared and fine-tuned among downstream models.} 
        \label{fig:diagram}
    \end{minipage}
    \hfill
    \begin{minipage}[b]{0.39\textwidth}
        \centering
        \includegraphics[width=\textwidth]{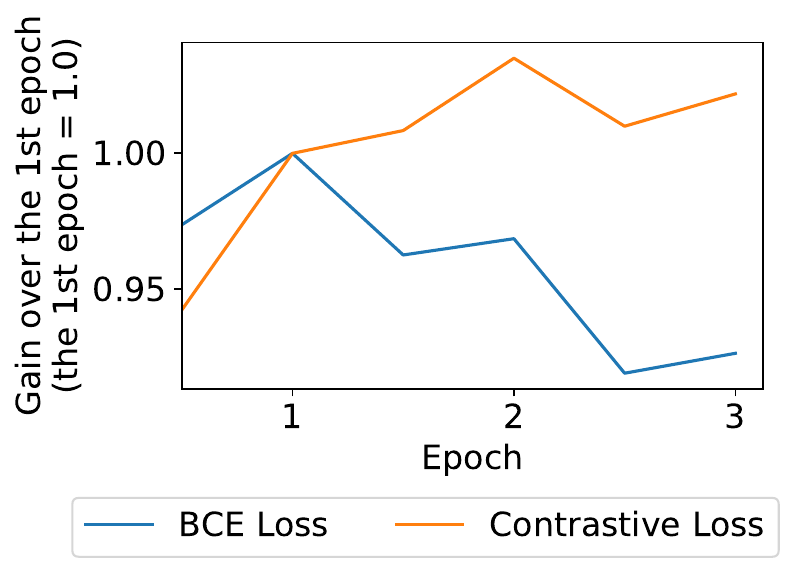}
        \vspace*{-1\baselineskip}
        \caption{In the pre-training stage, the binary entropy loss (BCE) overfits, but the contrastive loss generalizes well over epochs due to the reduced effective dimension of freedom.}
        \label{fig:epoch}
    \end{minipage}
    \vspace{-3mm}
\end{figure*}

\section{Design Choices}
Trade-offs are inevitable in industrial recommendation systems. Here, we discuss the key design choices in our embedding systems.

\vspace*{0.3\baselineskip}
\textbf{Trade-off between Model Complexity and Training Cost.} Due to the limitations of the one-epoch phenomenon, CTR models often require extended training windows to cover tail IDs and reach convergence, leading to substantial storage costs. To manage these costs, data down-sampling \cite{anil2022factory} is commonly employed to control the expenses of both model training and data storage. However, this can compromise ID coverage, resulting in suboptimal embeddings and reduced performance. Thus, we decided to separate the learning of ID-based embeddings from downstream models.

\vspace*{0.3\baselineskip}
\textbf{Two-stage Training.}  To address the one-epoch phenomenon, we separate the training of our ID embeddings from the downstream models. The first stage involves pre-training the shared foundational ID embeddings using a lightweight ID-only dataset and a minimal dot-product model. This approach ensures a robust starting point for embeddings and trade model complexity with extensive data coverage. These pre-trained embeddings are fine-tuned in a downstream recommendation model through multi-epoch training in the subsequent stage. This method allows for improved data utilization and multi-epoch learning, ultimately enhancing overall model performance and efficiency.

\section{Our Approach}
We propose pre-training foundational ID embeddings using a minimal dot-product model with contrastive loss, incorporating in-batch and uniform random negatives \cite{pancha2022pinnerformer}, as defined below:
\begin{equation}
       \mathcal{L}(u_i, p_i) = -\log \left( \nicefrac{e^{u_i^T p_i / \tau }}{\big(e^{u_i^T p_i / \tau } + \sum_{j=1}^{N} e^{u_i^T n_j / \tau }\big)}\right),
\end{equation}
where $u_i$ represents the user ID embedding, $p_i$ is the item ID embedding, $\tau$ is the learned temperature, and $\{n_j\}_{j=1,\ldots, N}$ are the in-batch and uniformly sampled random negatives.

Our experiments show that contrastive loss with combined negatives effectively mitigates the one-epoch overfitting phenomenon by reducing the effective dimensionality of tail entries. Additionally, the minimal and fast pre-training model allows us to enhance data coverage by using 10x more engagement data from various sources compared to downstream models. With improved generalization, the resulting ID embeddings are then integrated into downstream recommendation models and fine-tuned for specific tasks, as illustrated in Figure~\ref{fig:diagram}.

\section{Experiments and Results}
Here, we showed that the proposed contrastive loss function with combined negatives resists the one-epoch overfitting phenomenon. Then, we conducted ablation on the pre-training schema for ID embeddings and showed significant improvement over the baseline in multiple surfaces in online A/B experiments.

\textbf{Overfitting in the Pre-training Stage.} 
We compared the Hit$@3$ evaluation metric during the pre-training stage using traditional binary cross-entropy (BCE) loss and the proposed contrastive loss with negative samples. Figure~\ref{fig:epoch} illustrates that the proposed loss function successfully mitigates the one-epoch phenomenon, whereas BCE loss, commonly used in deep CTR models, exhibits overfitting patterns \cite{zhang2022towards} as expected.

\textbf{Two-stage vs Single-stage Pre-training.} In this experiment, we perform an ablation study with three methods: (1) merging the pre-training loss function into a single stage, (2) conducting two-stage pre-training but freezing the embedding in the downstream models, and (3) conducting two-stage pre-training and then training/fine-tuning the embedding with the downstream models.  All downstream models are trained with multi-epochs in three methods. As shown in Table~\ref{table:ablation_result}, single-stage training underperforms compared to the baseline due to overfitting. Moreover, fine-tuning the embedding in the downstream models consistently yields better performance than freezing the embedding.

\begin{table}[H]
\caption{Hit$@3$ performance across various training schema. Single-stage training is worse than the baseline due to overfitting, and two-stage strategies are always better. Additionally, fine-tuning performs better than freezing the embedding.}
\vspace*{-\baselineskip}
\begin{tabular}{ccc}
\toprule
\multirow{2}{*}{Pre-training Methods} & \multicolumn{2}{c}{Downstream Lift} \\
\cmidrule{2-3}
& Homefeed & Related Pins \\ \midrule
Baseline                & $+0.000\%\pm 0.032\%$ & $+0.000\%\pm 0.066\%$ \\
Single Stage            & $-3.347\%\pm 0.160\%$ & $-1.907\%\pm 0.116\%$ \\
Two-stage Frozen     & $+1.157\%\pm 0.025\%$ & $+1.929\%\pm 0.070\%$ \\
Two-stage Fine-tuned  & $\textbf{+1.323\%}\pm 0.048\%$ & $\textbf{+2.187\%}\pm 0.041\%$ \\
\bottomrule
\end{tabular}
\label{table:ablation_result}
\end{table}
\vspace*{-0.25\baselineskip}

\begin{table}[H]
\caption{Online A/B experiments with the two-stage approach in Homefeed and Related Pins ranking models show significant site-wide metrics gain for both use cases.}
\vspace*{-\baselineskip}
\begin{tabular}{ccc}
\toprule
\multirow{2}{*}{Online Metric} & \multicolumn{2}{c}{Downstream Lift} \\
\cmidrule{2-3}
& Homefeed & Related Pins \\ \midrule
Engagement & $+1.11\%$ & $+1.09\%$ \\
\bottomrule
\end{tabular}
\label{table:online_result}
\end{table}

\textbf{Online A/B Experiments in Live Web-scale Traffic.}
To verify the effectiveness of the proposed two-stage system, we conducted a two-week online A/B experiment comparing to the current production system across the Homefeed and Related Pins surfaces. The results in Table~\ref{table:online_result} show a total of 2.2\% site-wide engagement gains (aggregated from the two surfaces).

\section{CONCLUSION}
We introduce a novel approach to addressing the one-epoch phenomenon by pre-training foundational ID embeddings in a minimal model using contrastive loss. This allows for greater data coverage and utilization. Experimental results demonstrate that this two-stage learning strategy reduces overfitting and enhances generalization when fine-tuned in more complex downstream models.

\section*{Acknowlegement}
We extend our gratitude to the individuals who contributed to the development of our online recommendation model. We thank Haomiao Li, Travis Ebesu, and Huizhong Duan for their work on related-pin surface; Matthew Lawhon, Xue Xia, and Dhruvil Deven Badani for their contributions on the homefeed experiments; Howard Nguyen, Nazanin Farahpour, Yuelin Zhang, and Saurabh Vishwas Joshi for their efforts on the serving infrastructure; and Chuck Rosenberg, Bee-Chung Chen, and Nikil Pancha for their general contributions. Their expertise and dedication were crucial to the success of this project.

\bibliographystyle{ACM-Reference-Format}
\bibliography{recsys2024_arxiv}

\end{document}